\documentclass{aa}
\input psfig.sty
\usepackage{graphics,times}

\begin{document}

\thesaurus{08(02.12.1; 08.03.4; 08.05.2; 13.09.6; 08.09.2 MWC 349)}

\title{CO band emission from MWC\,349}

\subtitle{I. First overtone bands from a disk or from a wind?}

\author{M. Kraus\inst{1}
  \and
   E. Kr\"ugel\inst{1} 
  \and
   C. Thum\inst{2} 
  \and
   T. Geballe\inst{3}
   }

\offprints{M. Kraus}
\mail{mkraus@mpifr-bonn.mpg.de}

\institute{Max-Planck-Institut f\"ur Radioastronomie, Auf dem H\"ugel 69, 
           53121 Bonn, Germany
      \and 
           Institut de Radio Astronomie Millimetrique, F--38406 Saint Martin 
           d'H\`{e}res, France
      \and 
           Gemini Observatory, 670 North A'ohoku Place, University 
           Park, Hilo, Hawaii 96720, USA
      }

\date{Received; accepted }

\maketitle

\begin{abstract}

We observed the near infrared emission in the wavelength range 2.28--2.5\,$\mu$m
from the peculiar B[e]-star MWC\,349. The spectra contain besides
the strong IR continuum the first overtone CO bands and most of the hydrogen
recombination lines of the Pfund series, both in emission. We also modeled the
spectra. The Pfund lines have a gaussian profile with a FWHM of $\sim
100$\,km/s, and it turned out that their emission is in LTE and optically thin. 
To explain the CO bands, several scenarios were investigated. We found that the 
CO band heads are formed under LTE and that the gas must have a temperature of 
3500 to 4000\,K. The width of the $2\rightarrow 0$ band head indicates 
kinematical broadening of 50 to 60\,km/s. We can obtain fits to the measured 
spectra assuming that the CO gas has a column density of $5\cdot 
10^{20}$\,cm$^{-2}$ and is located either at the inner edge of the rotating 
circumstellar disk. In this case, the disk must have a bulge which partly blocks
the radiation so that the observer sees only a sector on the far side where the 
radial velocities are small. Or the CO emission originates in a wind with 
gaussian line profiles. Both fits are of equal quality and satisfactory. In a 
third alternative where the fit is less convincing, the CO emission is 
optically thin and comes from an extended Keplerian disk.


\end{abstract}

\section{Introduction}

After the discovery of CO band head emission in the BN object by Scoville
et al. (\cite{scoville}) a large number of further detections in other young stellar 
objects (YSO) followed (see e.g. Geballe \& Persson \cite{geballe}; Carr 
\cite{carr}; Chandler et al. \cite{chandler}; Greene \& Lada \cite{green}; 
Najita et al. \cite{najita}).

Several scenarios were discussed to explain the origin of the hot (2500--5000 K)
and dense ($n > 10^{11}$\,cm$^{-3}$) CO component (for some examples see Calvet et
al. \cite{calvet}; Martin \cite{martin}), but the most likely location is a neutral 
disk or wind (Carr \cite{carr}; Chandler, Carlstrom, \& Scoville \cite{chandler95}). 
The disk model is especially supported by high resolution spectroscopic observations 
of the CO $2\rightarrow 0$ band head. The shape of this band head shows for 
several YSO the kinematic signature of Keplerian rotation (Carr et al. 
\cite{carr93}; Carr \cite{carr95}; Najita et al. \cite{najita}) and is a powerful 
tracer for the existence of a circumstellar disk around young stellar objects.

Also the peculiar B[e]-star \object{MWC\,349} shows the first overtone 
CO bands in emission. These bands were observed first by Geballe \& Persson 
(\cite{geballe}) with a velocity resolution of about 460\,km/s. 

MWC\,349 is a binary system consisting of the main component MWC\,349A, classified as 
a B[e]-star, and the B0 III star companion, MWC\,349B, localized $2\arcsec$ west of 
MWC\,349A. We are only interested in the main component, in the following refered to as 
MWC\,349. Its evolutionary state is still unclear. It shows some characteristics of a pre-main
sequence B[e]-type star as well as characteristics of a B[e] supergiant (e.g. Lamers et al.  
\cite{lamers}).

Cohen et al. (\cite{cohen}) determined its distance (1.2\,kpc), bolometric 
luminosity ($\sim 3\cdot 10^{4}$\,L$_{\sun}$) and the visual extinction towards it 
($A_{\rm V}^{\rm ISM}\simeq 10$\,mag of which 2\,mag might be circumstellar). 

The existence of a bulge of circumstellar dust around MWC\,349 has been known since long
ago (e.g. Geisel \cite{geisel}). The proposition that MWC\,349 also has a disk 
was supported by observations of double-peaked emission lines (Hamann \& Simon 
\cite{hamann}, \cite{hamann88}) and by IR speckle interferometry (Leinert 
\cite{leinert}; Mariotti et al. \cite{mariotti}) which revealed a disk-like structure
of the dust emission in the east-west direction, seen nearly edge on. An accumulation
of neutral gas and dust in the equatorial plane of the star is also expected to be 
responsible for the bipolar structure of the optically thick wind zone seen in the 
VLA-map of White \& Becker (\cite{white}). The mass loss rate found for a 50\,km/s
wind velocity is $\sim 1.2\cdot 10^{-5}$\,M$_{\sun}$yr$^{-1}$ (Cohen et al.
\cite{cohen}).

Another indicator for the circumstellar disk are the strong hydrogen recombination 
maser lines in the mm and submm range which also show the characteristic
double-peaked profiles (e.g. Mart\'{\i}n-Pintado et al. \cite{martin-pintado}). 
Recombination lines at different wavelengths sample different regions. With
decreasing quantum number $n$, i.e. increasing frequency, one sees ionized gas 
closer to the star. The fact that the rotational velocity of recombination lines 
with decreasing $n$ displays a systematical increase led to the assumption that the
double-peaked maser emission comes from the ionized atmosphere of a Keplerian 
rotating disk around a 25--30\,M$_{\sun}$ star (Thum, Mart\'{\i}n-Pintado \& 
Bachiller \cite{thum}; Thum et al. \cite{thum94}) whose free-free and free-bound
emission was modelled by Kraus et al. (\cite{kraus}). In addition, Rodr\'{\i}guez \& 
Bastian (\cite{rodriguez}) determined the inclination angle of the disk towards the line 
of sight to $15^{\circ}\pm 5^{\circ}$. 

In this paper, we present new low resolution observations of the total first overtone 
CO band emission as well as high resolution observations of the $2\rightarrow 0$ and
$3\rightarrow 1$ band heads. We discuss the probable location of the hot CO gas
by modeling the emission for several scenarios.

\section{Observations}\label{obs}

\begin{figure}[h!]
\resizebox{\hsize}{!}{\includegraphics{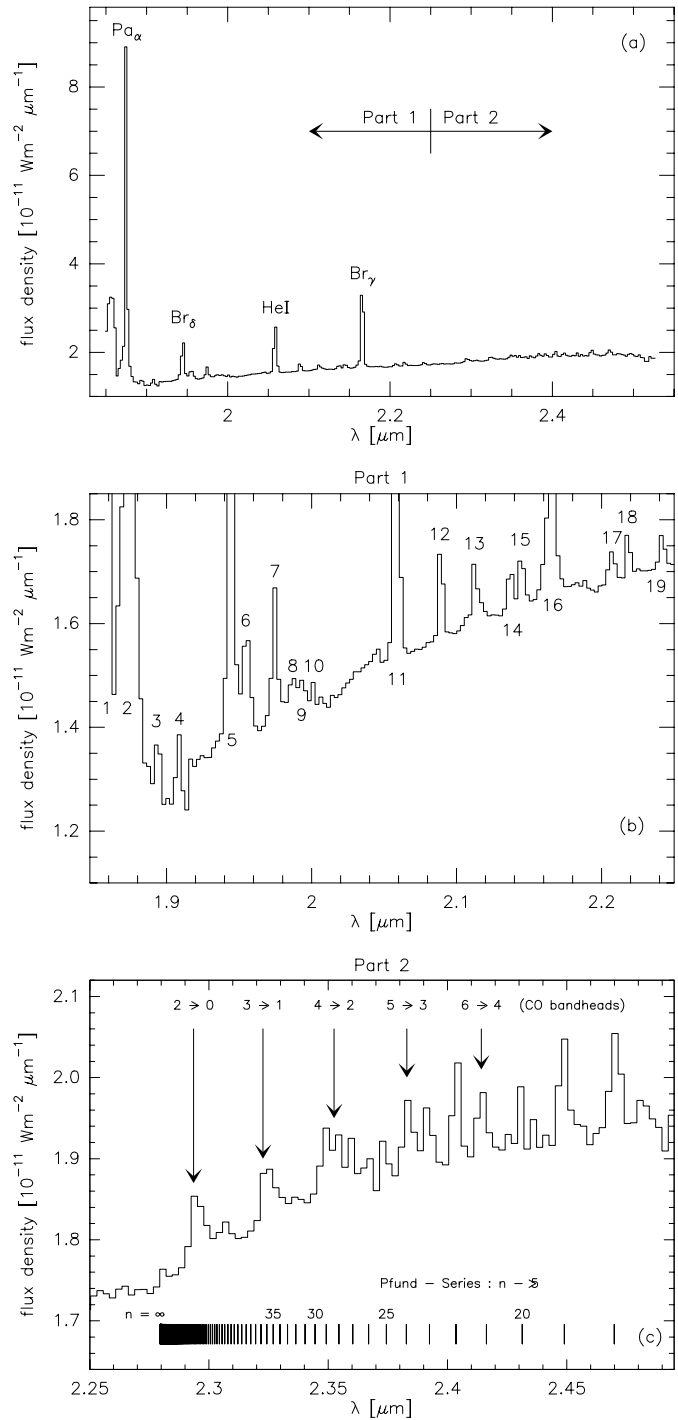}}
\caption{For better visualization, the total NIR spectrum of MWC\,349 observed with UKIRT with 
a spectral resolution of $\sim 330$\,km/s (upper panel) is split in two parts: Part 2 (bottom 
panel) contains the first overtone CO bands (indicated by the arrows) and most lines from the 
Pfund series of the hydrogen atom (vertical bars), the remaining spectrum (Part 1, 
middle panel) contains several emission lines listed in Table \protect{\ref{identi}}.}
\label{observ}
\end{figure}

In 1997 we observed MWC\,349 with the UKIRT telescope in the NIR range ($1.85 - 2.5\,\mu$m)
with a spectral resolution of $\sim 330$\,km/s (Fig.\,\ref{observ}a). The spectrum contains
several prominent hydrogen and helium recombination lines as well as the first overtone CO 
bands, all of them in emission.

\begin{table}[htb]
\caption{Identified lines in the NIR spectrum of MWC\,349
(Figs. \protect{\ref{observ}}a and b).}
\begin{tabular}{rrrcc}
\hline
\multicolumn{3}{c}{\textrm{number}} & \multicolumn{1}{c}{\textrm{wavelength [$\mu$m]}}
 & \multicolumn{1}{c}{\textrm{line identification}} \\
\hline
& 1 & & 1.86904 & \ion{He}{i} \\
& 2 & & 1.87561 & Pa$\alpha$ \\
& 3 & & 1.89282 & \ion{C}{i}, \ion{Ca}{i} \\
& 4 & & 1.90867 & \ion{He}{i} \\
& 5 & & 1.94508 & Br$\delta$, \ion{Ca}{i} \\
& 6 & & 1.95445 & \ion{N}{i} \\
& 7 & & 1.97470  & \ion{Si}{i}, \ion{C}{i}, \ion{N}{i} \\
& 8 & & 1.98702  & \ion{Ca}{i}, \ion{C}{i} \\
& 9 & & 1.99318  & \ion{Ca}{i}, \ion{Si}{i}, \ion{C}{i} \\
& 10 & & 2.00130  & \ion{C}{i} \\
& 11 & & 2.05917 & \ion{He}{i} \\
& 12 & & 2.08819 & (?)\,\ion{Sn}{i}, \ion{Fe}{ii} \\
& 13 & & 2.11194 & \ion{He}{i} \\
& 14 & & 2.13831 & \ion{Mg}{ii}, \ion{C}{i}, \ion{Si}{i} \\
& 15 & & 2.14359 & \ion{Mg}{ii}, \ion{Ca}{ii}, [\ion{Fe}{iii}] \\
& 16 & & 2.16611 & Br$\gamma$ \\
& 17 & & 2.20687 & \ion{Na}{i} doublet, \ion{Si}{i}, (?)\,\ion{O}{ii} \\
& 18 & & 2.21741 & \ion{C}{i} \\
& 19 & & 2.24113 & (?)\,\ion{Fe}{ii} \\
\hline
\end{tabular}
\label{identi}
\end{table}

\begin{figure}[htb]
\resizebox{\hsize}{!}{\includegraphics{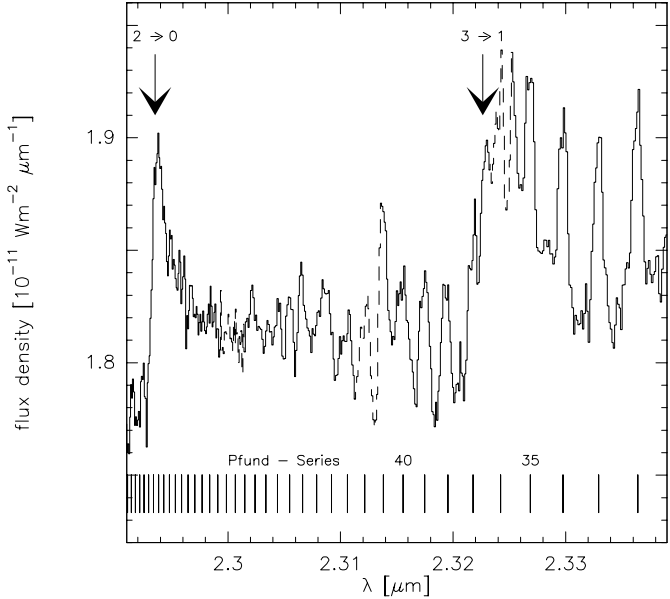}}
\caption{Resulting calibrated high resolution spectrum of the first 
two CO band heads from MWC\,349. The dashed regions mark the overlap for each 
two subspectra where the calibration is rather uncertain.}
\label{hires}
\end{figure}

In two sessions in 1998 we reobserved MWC\,349 in the spectral range 
2.285--2.341\,$\mu$m with higher spectral resolution (10--15\,km/s). The 
observations consist  of four seperate subspectra with a small overlap at their 
edges. They are slightly offset relative to each other and at their low 
wavelength end, the fluxes are systematically overestimated. We calibrated the 
subspectra by smoothing them to the resolution of the spectrum in 
Fig.\,\ref{observ}c, then scaled each spectrum with a factor near $\sim 1$ and 
combined them to the final plot of Figure \ref{hires}.

Unfortunately, the $3\rightarrow 1$ band head falls into the overlap region of two
subspectra, so its strength is rather uncertain (marked by the dashed line within the
figure). This fact turned out during the modeling to be a disadvantage.

\section{Results}

\subsection{Hydrogen recombination lines -- the Pfund series}\label{pfund}

As can be seen in Figs.\,\ref{observ}c and \ref{hires}, the spectrum of MWC\,349 in 
the wavelength range from 2.25 to 2.5\,$\mu$m contains beside the first overtone CO 
bands also the Pfund series (transitions of the form $n \rightarrow n' = 5$, denoted as 
Pf($n$)\,) of the hydrogen atom. Hamann \& Simon (\cite{hamann}) already identified some
of them in their velocity-resolved infrared spectroscopy of MWC\,349, but their 
resolution of $\sim$20\,km/s was lower than ours, so they could only resolve lines up to 
Pf(34). Their FWHM values of several unsmeared lines were about 100\,km/s and agree with 
ours.

Many of the other emission lines they observed are double-peaked which is a hint for
a rotating medium. At first glance we cannot exclude the Pfund lines to take part of the 
rotation because they are superimposed on the CO bands, but their profile looks more than
a gaussian than a double-peaked one. A gaussian velocity component of about 
50\,km/s, however, is not a surprising feature for MWC\,349. Also the cm and mid IR 
hydrogen recombination lines as well as the 'pedestal` feature of the double-peaked 
mm recombination maser lines show such a velocity component (Smith et al. 
\cite{smith}; Thum, Mart\'{\i}n-Pintado \& Bachiller \cite{thum}). It is often 
ascribed to the wind of MWC\,349.

To model the hydrogen emission, we assume the Pfund lines to be optically thin. 
Their intensity is given by 
\begin{equation}\label{intens}
I_{\nu} =  N_{n} A_{n,5} h\nu~\Phi_{\rm H}(\nu)~. 
\end{equation}
To evaluate it, we need the number density $N_{n}$ of the atoms in level $n$, the 
Einstein coefficients, $A_{n,5}$, and the profile function of the hydrogen gas, 
$\Phi_{\rm H}(\nu)$.

In view of the measured line widths, we choose a gaussian profile function
defined as
\begin{equation}\label{gaussprofil}
\Phi_{\rm H} (\nu) = \frac{1}{\sqrt{\pi}\frac{\nu_{0}}{c}v}\,\exp\left[-\left(\frac{\nu - 
\nu_{0}}{\frac{\nu_{0}}{c}v}\right)^{2}\right]
\end{equation}
$v$ is the most probable velocity which is in the case of the thermal velocity given by 
$v=\sqrt{2kT/m}$ for particles having a Maxwellian velocity distribution. In our case, $v\simeq 
$\,50\,km/s. We also take into account the local standard of rest velocity of $\sim$\,8\,km/s 
(Thum, Mart\'{\i}n-Pintado \& Bachiller \cite{thum}; Thum et al. \cite{thum95}).

\begin{figure}[htb]
\resizebox{\hsize}{!}{\includegraphics{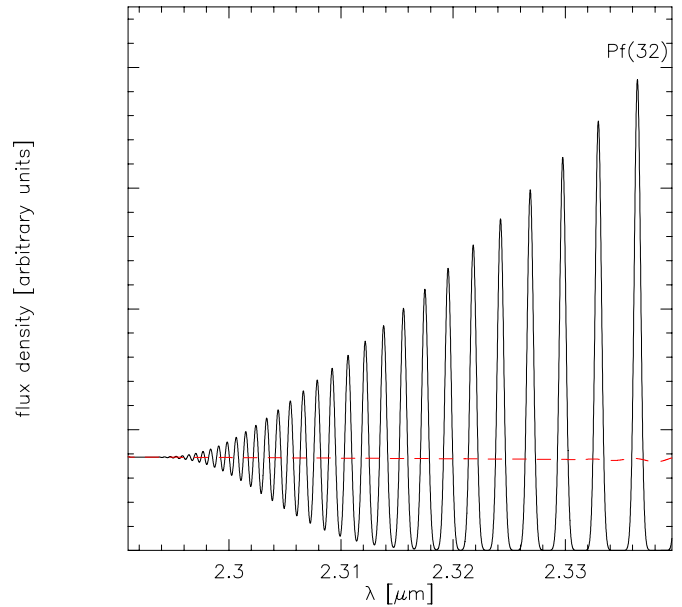}}
\caption{Spectrum of optically thin Pfund lines in LTE, with an isothermal electron
temperature of $T_{\rm e} = 10^{4}$\,K and a gaussian velocity component of 50\,km/s.
The spectral resolution to which the lines were smoothed are 330\,km/s (dashed line)
and 15\,km/s (solid line). At short wavelengths, the lines blend into a 'continuum`,
especially for the low resolution.}
\label{pf_syn1}
\end{figure}

The Einstein coefficients, $A_{nn'}$, can be calculated, following Menzel \& Pekeris
(\cite{menzel}). If we restrict ourselves to Pfund lines with $n\geq 18$ 
(see Fig.\,\ref{observ}c), they can be approximated by 

\begin{equation}\label{a_n5}
A_{n,5} \simeq 3.023\cdot 10^{9}~n^{-5.005}~.
\end{equation}

\begin{figure}[htb]
\resizebox{\hsize}{!}{\includegraphics{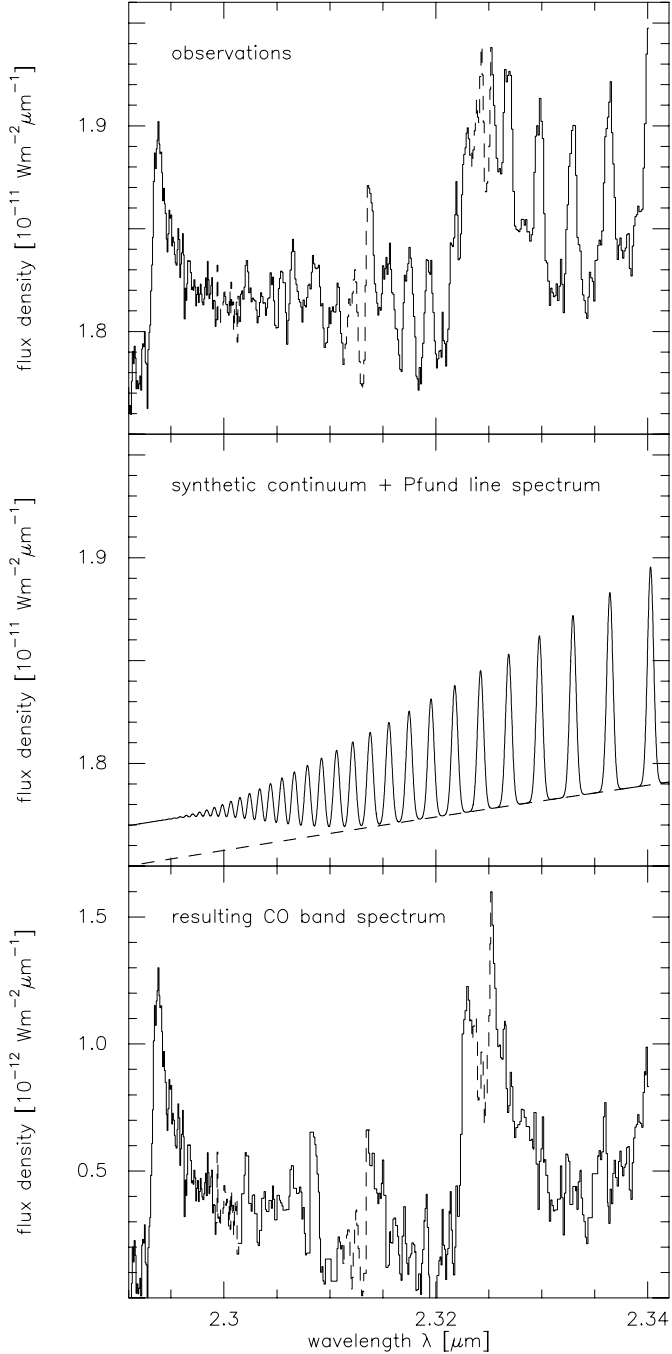}}
\caption{Subtraction of the synthetic continuum and Pfund line emission (solid line,
middle panel, pure continuum is shown by the dashed line) from the observations (upper 
panel) leads to the almost pure CO first overtone band emission (lower panel).}
\label{pf_subtract}
\end{figure}

The number density, $N_{n}$, of the level $n$ follows from Saha's equation
\begin{equation}\label{bes_n}
N_{n} = b_{n} N_{\rm i} N_{\rm e} g_{n}~\frac{h^{3}}{2\,(2\pi
m_{\rm e}kT_{\rm e})^{3/2}}~{\rm e}^{\frac{h\,Ry}{n^{2}kT_{\rm e}}}
\end{equation}
where $N_{\rm i}$ and $N_{\rm e}$ are the number densities of the protons and electrons,
$T_{\rm e}$ and $m_{\rm e}$ are the electron temperature and electron mass, $g_{n}=2n^{2}$ is
the statistical weight of the state $n$, and $Ry$ is the Rydberg constant. The
$b_{n}$-factors which describe the deviation from LTE and which depend on the electron
temperature and density were taken from the electronic Tables of Storey \& Hummer
(\cite{storey}) for Menzel's case B recombination theory.

Detailed modeling of the Pfund line spectrum performed by Kraus 
(\cite{kraus00}) showed that the emission arises within the innermost parts of 
the ionized wind where the electron densities are highest. Within these regions
it is expected that $N_{\rm e}\leq 10^{9}$\,cm$^{-3}$. Electron densities as high 
as $10^{8}$\,cm$^{-3}$ have already been found by Strelnitski et 
al.\,(\cite{strelnitski96}) and Thum et al.\,(\cite{thum98}). For 
such high electron densities the emission is clearly in LTE, i.e. the
$b_{n}$-factors are $\sim\,1$. Therefore, non LTE effects are negligible. 

In our case of optically thin Pfund lines in LTE and with a constant electron 
temperature the intensity ratios of two neighbouring Pfund lines is according to
Eq.\,(\ref{intens}) independent of the electron density
\begin{equation}
\frac{I_{{\rm Pf}(n+1)}}{I_{{\rm Pf}(n)}}\sim \left( \frac{n}{n+1}\right)^{5.005}
\cdot \exp\left(-\frac{hRy}{kT_{\rm e}}\frac{2n+1}{(n^{2}+n)^{2}}\right)~.
\end{equation}
and always smaller than 1, which means that the intensity of the Pfund lines decreases
with increasing quantum number. This behaviour can qualitatively be seen in 
Fig.\,\ref{pf_syn1} where we plotted the synthetic Pfund lines which we smoothed to the 
spectral resolution of 10--15\,km/s (solid line) and 330\,km/s (dashed line),
respectively. The wavelength difference between the neighbouring Pfund lines decreases 
with increasing $n$ leading to a blend of the individual lines. The onset of this blend 
depends on the width of the lines and on the spectral resolution and leads to a hydrogen
'continuum`. 

\begin{table}[htb]
\caption{Parameters for hydrogen used for the calculations of the Pfund lines shown in
Fig.\,{\protect\ref{pf_subtract}}\,(middle panel).}
\begin{tabular}{cccc}
\hline
$T_{\rm e}$ [K] & $v_{\rm gauss}$ [km/s] & $A^{\rm ISM}_{\rm v}$ [mag] & $\int N_{\rm e}^{2}dV$
 [cm$^{-3}$] \\
\hline
$\leq 10^{4}$ & $\sim 50$ & 10 & $6.35\cdot 10^{60}$ \\
\hline
\end{tabular}
\label{pf_param}
\end{table}

The interstellar visual extinction towards MWC\,349 is about $A_{\rm v}^{\rm ISM} \simeq 
10$\,mag and has to be taken into account. Fitting the Pfund lines with the parameters 
displayed in Table \ref{pf_param} leads to the quantity $\int N_{\rm e}^{2}dV$, where $V$ is 
the volume of the emitting region. 

The middle panel of Figure \ref{pf_subtract} contains the modelled Pfund series added
to the continuum emission of MWC\,349 (dashed line) taken from Kraus et al. 
(\cite{kraus}). These two components were subtracted from the observations (upper panel) 
and the resulting spectrum (lower panel) can mainly be ascribed to the CO first overtone
band emission. To fit this observed CO spectrum is the aim of the following sections.

\subsection{Theory of the CO bands}

The energy of a diatomic molecule in rotational level $J$ and vibrational level $v$ can be
expanded in the following way (Dunham \cite{dunham32a}, \cite{dunham32b}) 
\begin{equation}
E(v,J) = hc~\sum_{k,l} Y_{k,l}\,\left( v+\frac{1}{2}\right)^{k}(J^{2}+J)^{l}~.
\end{equation}
The parameters $Y_{k,l}$ are for the CO molecule taken from Farrenq et al. 
(\cite{farrenq}). The first overtone bands result from coupled vib-rot transitions in 
the ground electronic state and obey the selection rules $\Delta v = 2$ and $\Delta J = 
\pm 1$.

In the following, we assume the CO gas to be in LTE. Then, the levels are populated 
following a Boltzmann distribution,
\begin{equation}\label{n_vib_rot}
N_{vJ} = \frac{N}{Z} (2J + 1)\,e^{\frac{-E(v,J)}{kT}}
\end{equation}
where $N$ and $T$ are the total number density and temperature of CO molecules and 
$Z$ is the total partition function, given as the product of the vibrational and 
the rotational partition function 

\begin{equation}
Z = Z_{v}\cdot Z_{J} = \sum_{v} e^{\frac{-E_{v}}{kT}} \cdot \sum_{J} 
(2J + 1)\,e^{\frac{-E_{J}}{kT}}~.
\end{equation}

To account for optical depth effects, we calculate the line intensities from the 
transfer equation 
\begin{equation}\label{line_intens}
I_{\nu} = B_{\nu}(T)\,(1-e^{-\tau_{\nu}})\,.
\end{equation}
The optical depth is $\tau_{\nu} = \int\kappa_{\nu}\,ds$ with the absorption coefficient
per cm
\begin{equation}\label{kappa}
\kappa_{\nu} = \frac{c^{2}N_{vJ} A_{vJ;v'J'}}{8\pi \nu^{2}}\,\left( \frac{2j + 1}
{2j' + 1}\cdot\frac{N_{v'J'}}{N_{vJ}} - 1\right)\,\Phi_{\rm CO}(\nu)
\end{equation}
where $\Phi_{\rm CO}(\nu)$ is the profile function of the CO gas. The Einstein 
coefficients, $A_{vJ;v'J'}$, are from Chandra, Maheshwari \& Sharma (\cite{chandra}). 

\begin{figure}[htb]
\resizebox{\hsize}{!}{\includegraphics{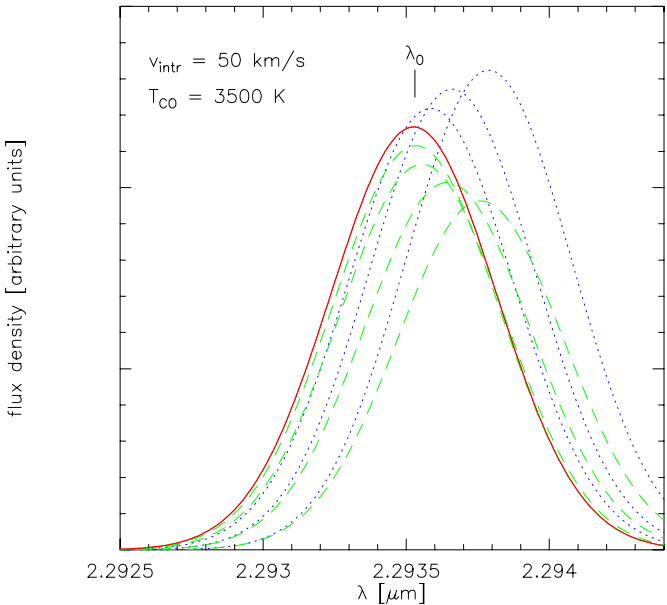}}
\caption{Individual vib-rot lines that form the $2\rightarrow 0$ band head.
The solid curve is the $(2,51) \rightarrow (0,50)$ transition. The dotted curves are the
neighbouring transitions with rotational quantum numbers decreasing in steps of one; for 
the dashed curves, the quantum numbers increase in steps of one.}
\label{line}
\end{figure}

\begin{figure}[htb]
\resizebox{\hsize}{!}{\includegraphics{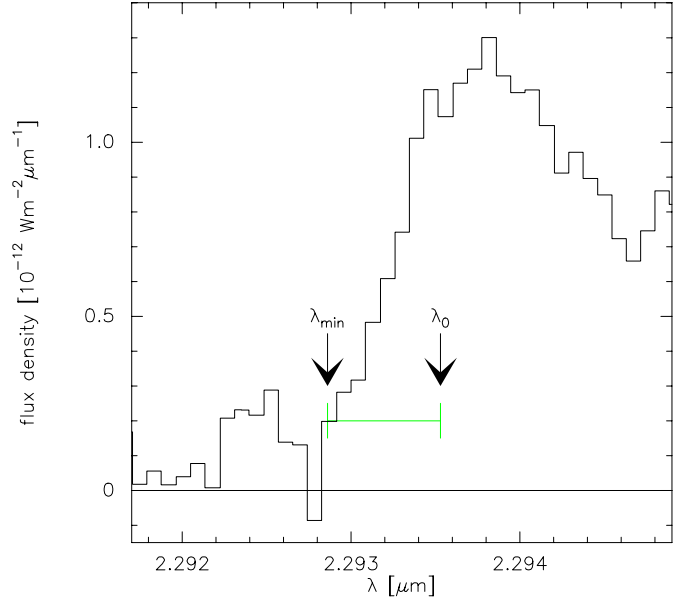}}
\caption{The onset of the spectrum above noise is choosen as the minimum wavelength
$\lambda_{\rm min}$.}
\label{mini}
\end{figure}

Let $\lambda_{\rm min}$ denote the onset of the ($2\rightarrow 0$) bandhead.  We define
it to be the wavelength where the intensity of the $(2,51)\rightarrow (0,50)$ line (see
Fig.\,\ref{line}) has dropped by a factor $x \sim 5$ from its maximum at $\lambda_0$; the
exact number is not important.  We determine $\lambda_{\rm min}$ from Figure
\ref{mini} to $(2.29285\pm 0.00005)\,\mu$m.

The value of $\lambda_{\rm min}$ depends of course on the various mechanisms of
line broadening.  Most processes lead to a gaussian or nearly gaussian profile,
like thermal (turbulent) motion of mean velocity $v_{\rm th}$ ($v_{\rm turb}$),
detector resolution $v_{\rm res}$, spherical wind $v_{\rm wind}$ (see e.g. Hamann \&
Simon \cite{hamann}).  Their superposition results again in a gaussian line of width
$v_{\rm gauss} = \sqrt{v_{\rm th}^{2} + v_{\rm turb}^{2} + v_{\rm wind}^{2} +
v_{\rm res}^{2}}$.

Disk rotation, on the other hand, produces a double-peaked profile.  In a gas
ring rotating at velocity $v_{\rm rot}$ a line with laboratory frequency $\nu_0$ is
shifted up to a maximum frequency $\nu_0\cdot(1+v_{\rm rot}/c)$.  In a mixture
of disk rotation and gaussian broadening one finds the following expression for
the minimum wavelength,
\begin{equation}\label{l_min}
\frac{1}{\lambda_{\rm min}} = \frac{1}{\lambda_{0}}\left[ 1+\frac{v_{\rm lsr} +
v_{\rm rot}}{c}\right] \left[\frac{v_{\rm gauss}}{c}\,\sqrt{\ln x} + 1\right]\,,
\end{equation}
where $v_{\rm lsr} = 8$\,km/s is the velocity shift of MWC\,349 with respect to
the local standard of rest.  This relation only holds if the emission in the rise 
of the spectrum is optically thin. The thermal velocity $v_{\rm th}$ of CO molecules
at 3500\,K is less than $\sim$\,1\,km/s and negligible in comparison with the
spectral resolution $v_{\rm res} = 10\ldots 15$\,km/s.  The same can be said
about the turbulent velocity which is not likely to exceed the speed of sound.

\begin{figure}[htb]
\resizebox{\hsize}{!}{\includegraphics{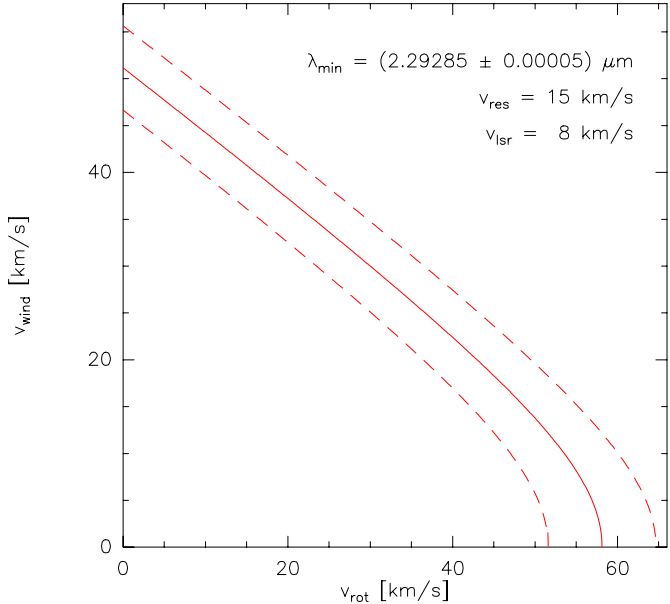}}
\caption{The dependence of $v_{\rm wind}$ on $v_{\rm rot}$ after
Eq.\,({\protect{\ref{l_min}}}) is shown. $x=10$, and the remaining parameters
are chosen as indicated in the Figure. The solid line is for $\lambda_{\rm min} =
2.29285\,\mu$m, the dashed line above (below) is for the lower (upper) limit of
$\lambda_{\rm min}$.}
\label{v_turb}
\end{figure}

Eq.\,(\ref{l_min}) can now be used to relate $v_{\rm wind}$ to the rotational
velocity $v_{\rm rot}$.  The result is shown in Figure \ref{v_turb}. The dashed
lines above and below the solid line are for the lower and upper limit of
$\lambda_{\rm min}$.  If the CO gas performs no rotation, the gas must emanate
from the disk in a wind with about $\sim$50\,km/s, the same velocity as for the
Pfund lines.  Without a wind, one derives a rotational velocity of $\sim 58\pm
7$\,km/s.  This value is larger than other rotational velocities around MWC\,349
observed up to now, the highest being 42.5\,km/s for \ion{He}{i} (Hamann \&
Simon \cite{hamann}).

\section{Modeling the CO band emission}\label{model}

\subsection{A gaussian line profile for the CO gas}\label{turb_mot}

First, we neglect rotation and assume a pure gaussian line profile. The velocity 
of the CO gas is taken from Figure\,\ref{v_turb} as $v_{\rm gauss} \approx 50$\,km/s. 

\begin{figure}[htb]
\resizebox{\hsize}{!}{\includegraphics{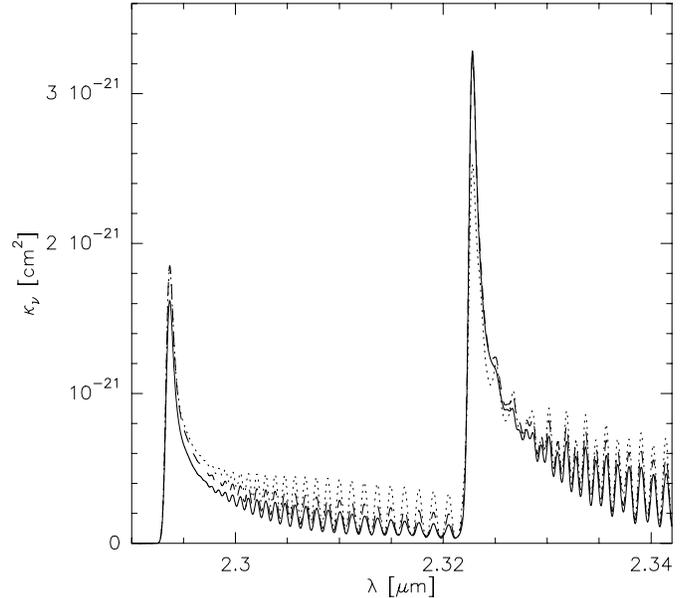}}
\caption{Absorption coefficient per CO molecule, $\kappa_{\nu}$, for temperatures of
3000\,K (dotted line), 4000\,K (dashed line), and 5000\,K (solid line). The gaussian
velocity of the CO gas is 50\,km/s.}
\label{kappa_turb}
\end{figure}

For the calculation of the optical depth we assume the absorption coefficient to be
constant along the line of sight. Then $\tau_{\nu}$ is simply given as the product of 
the absorption coefficient per CO molecule, $\kappa_{\nu}$, times the CO column density,
$\tau_{\nu} = \kappa_{\nu} \cdot N_{\rm CO}$. 

\begin{figure}[htb]
\resizebox{\hsize}{!}{\includegraphics{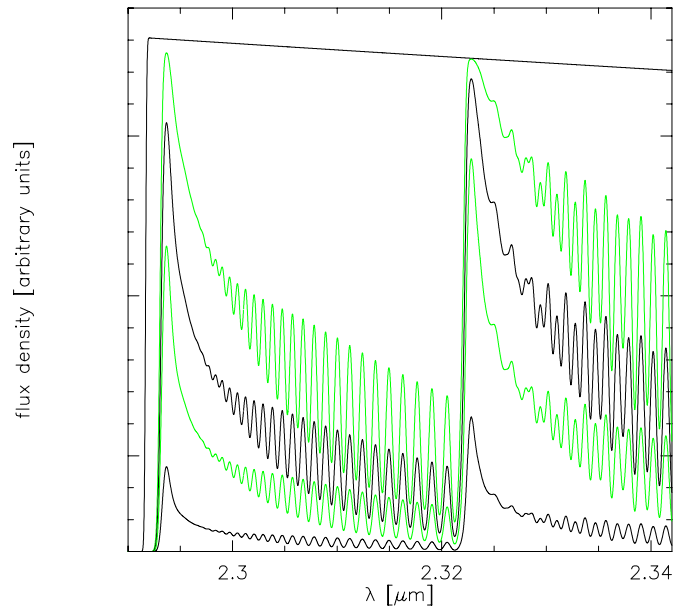}}
\caption{Calculated CO band emission spectra for a temperature of the CO gas of 
4000\,K, a gaussian velocity of $\sim 50$\,km/s, and a CO column density increasing 
from bottom to top: $10^{20}, 5\cdot 10^{20}, 10^{21}$, and $2\cdot 10^{21}$\,cm$^{-2}$.
The upper line represents the case for $\tau\rightarrow\infty$.} 
\label{turb_syn}
\end{figure}

In Fig.\,\ref{kappa_turb} we plotted $\kappa_{\nu}$ for temperatures between 3000\,K and
5000\,K, and a gaussian velocity of the gas of 50\,km/s. In the wavelength range around
the $3\rightarrow 1$ band head (2.322--2.325\,$\mu$m) the absorption coefficient is higher
than in the region around the $2\rightarrow 0$ band head (2.293--2.295\,$\mu$m). This
effect becomes stronger with increasing temperature. Thus, the turnover from optically
thin to optically thick emission depends not only on the column density but also on the
temperature: the $3\rightarrow 1$ band head
becomes optically thick at lower temperature and lower column density than the $2
\rightarrow 0$ band head.

\begin{figure}[htb]
\resizebox{\hsize}{!}{\includegraphics{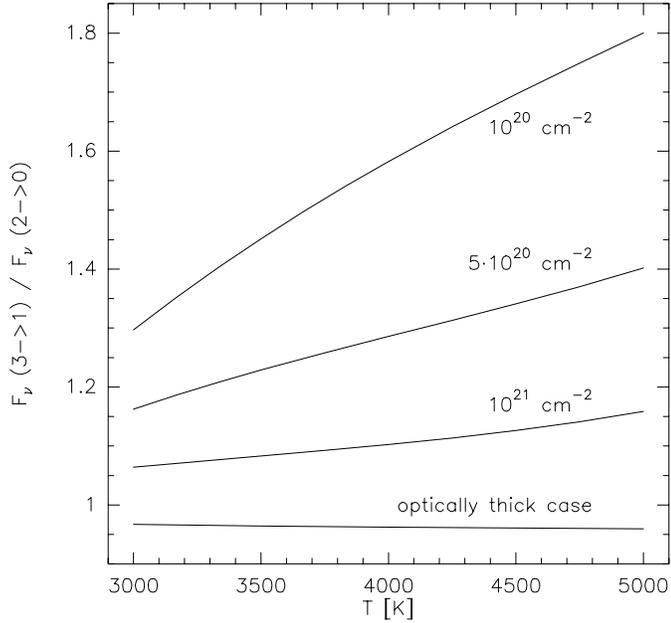}}
\caption{The flux density ratio of the $3\rightarrow 1$ and $2\rightarrow 0$ band heads
as a function of temperature and column density. The ratio is highest in the optically 
thin case ($N_{\rm CO} \leq 10^{20}$\,cm$^{-2}$) and approaches $\simeq 0.96$ for $\tau
\rightarrow\infty$.}
\label{verhaelt}
\end{figure}

\begin{figure}[htb]
\resizebox{\hsize}{!}{\includegraphics{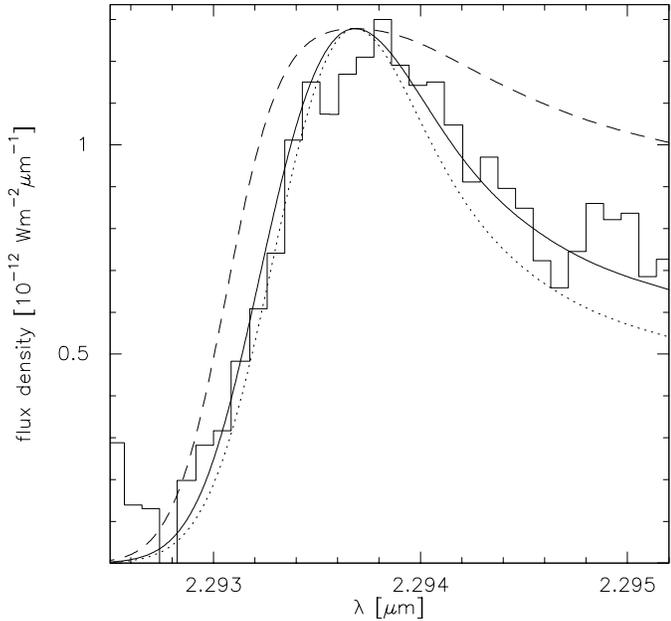}}
\caption{Theoretical ($2\rightarrow 0$) CO band head emission for a gaussian line profile
with $v_{\rm gauss} \sim 50$\,km/s. The CO column density is $2\cdot 10^{21}$\,cm$^{-2}$
(dashed line), $5\cdot 10^{20}$\,cm$^{-2}$ (solid line) and $10^{20}$\,cm$^{-2}$ (dotted
line).}
\label{turb_tief}
\end{figure}

Next, we vary the CO column density and fix $T_{\rm CO}$ at 4000\,K. The synthetic 
spectra in Figure \ref{turb_syn} are calculated for the case of extreme optical depth
(upper line) and for CO column densities (from top to bottom) of $2\cdot 
10^{21}, 10^{21}, 5\cdot 10^{20}$, and $10^{20}$\,cm$^{-2}$. With increasing column 
density the optical depth enlarges and the intensity tends towards its limiting blackbody 
value. In addition, the spectra show two characteristic features: a varying intensity 
ratio of the two band heads and a broadening of the band head structure. 

The first property can better be seen in Figure \ref{verhaelt} where we plotted the flux 
density ratio of the $3\rightarrow 1$ and $2\rightarrow 0$ band head with temperature for
different column densities. This ratio decreases with increasing column density, i.e.
increasing $\tau$, and approaches 0.96 for $\tau\rightarrow\infty$. 

\begin{figure}[htb]
\resizebox{\hsize}{!}{\includegraphics{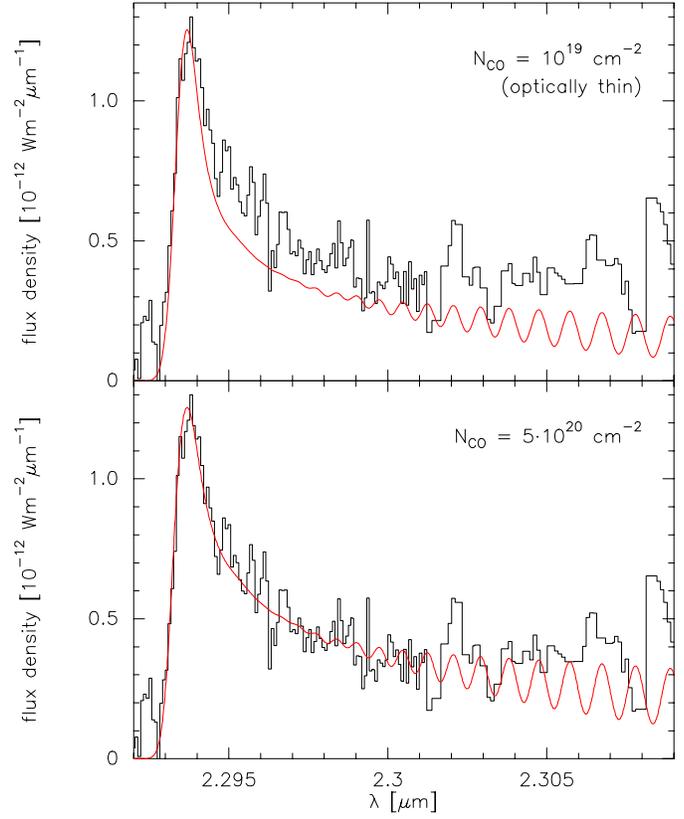}}
\caption{Comparison between theoretical ($2\rightarrow 0$) band head spectra of optically
thin (upper panel) and marginally optically thick emission (lower panel). In both cases,
the temperature is 3500\,K.}
\label{thick_thin}
\end{figure}

The determination of the column density with only the help of this ratio is, however, 
not definite because the flux density ratio also depends on temperature. From our 
observations we get no information at all on the column density in this way because the
observed strength of the $3\rightarrow 1$ band head is subject to some uncertainty as 
discussed in Section \ref{obs}. 

The second characteristic feature is the broadening of the band head structure with 
increasing column density, especially at the long wavelength side. In Figure 
\ref{turb_tief} the observed $2\rightarrow 0$ band head region (histogram) is overlaid 
on the theoretical spectra with column densities of $2\cdot 10^{21}$ (dashed 
line), $5\cdot 10^{20}$ (solid line), and $10^{20}$\,cm$^{-2}$ (optically thin case,
dotted line).
 
At first glance, it seems that the observations might also be fitted by optically thin 
emission, but the upper panel of Figure \ref{thick_thin} which includes a somewhat larger
wavelength interval is clearly inconsistent with the optically thin case whereas the
observations can be reconciled with a synthetic spectrum of column density $N_{\rm CO}
\simeq 5\cdot 10^{20}$\,cm$^{-2}$ (lower panel of Figure \ref{thick_thin}). An additional
disagreement is that the flux density ratio of the bandheads is highest for the optical 
thin case which can not be confirmed by the observations. 
  
\begin{figure}[htb]
\resizebox{\hsize}{!}{\includegraphics{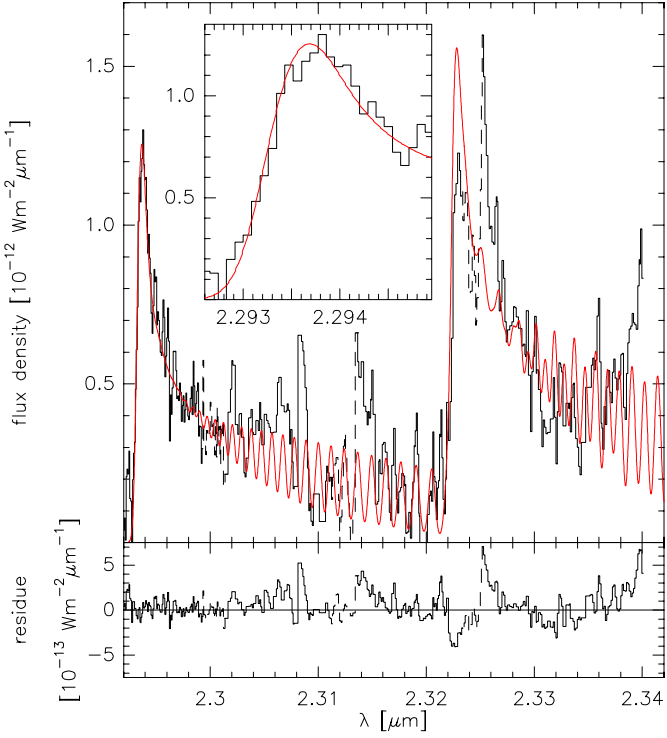}}
\caption{Modelled CO band emission (black line) for a temperature of $T_{\rm CO}=
3500$\,K and a gaussian velocity of $\sim 50$\,km/s overlaid on the observations
(grey histogram). The model parameters are given in Table {\protect{\ref{param_turb}}}.
The insert shows a blow-up of the $2\rightarrow 0$ band head. Lower panel: residue
of observations minus model. Some yet unidentified emission lines seem to be present.}
\label{turb}
\end{figure}

So we take $N_{\rm CO} = 5\cdot 10^{20}$\,cm$^{-2}$ as the best fit CO column density.
The synthetic spectra of the $2\rightarrow 0$ band head look very similar for
temperatures in the range of $3000 - 5000$\,K. For a more sensitive limitation of the
temperature range we must take a look at the $3\rightarrow 1$ band head. From there we
conclude that $T_{\rm CO} =$\,3500--4000\,K. Unfortunately, an exact determination of
the CO temperature fails because the higher bandheads are not available within our
high resolution spectrum.  

\begin{table}[htb]
\caption{Parameters for CO used for the calculations shown in Fig.\,{\protect
\ref{turb}}. $A_{\rm v}^{\rm ISM}$ denotes the foreground extinction, $A_{\rm CO}$ and
$M_{\rm CO}$ are the emitting CO area projected to the sky and the mass of CO derived from
the fit, respectively.}
\begin{tabular}{cccc|cc}
\hline
$N_{\rm CO}$ & $T_{\rm CO}$ & $v_{\rm gauss}$ & $A^{\rm ISM}_{\rm v}$ & $A_{\rm CO}$ &
$M_{\rm CO}$ \\
$[{\rm cm}^{-2}$] & [K] & [km/s] & [mag] & [cm$^{2}$] & [g] \\
\hline
$5\cdot 10^{20}$ & 3500 & $\sim 50$ & 10 & $1.36\cdot 10^{26}$ & $3.2\cdot 10^{24}$ \\
\hline
\end{tabular}
\label{param_turb}
\end{table}

With the derived values for column density and temperature we model the synthetic CO band
spectrum taking into account the foreground extinction, $A^{\rm ISM}_{\rm v}$, and fitted it
to the observations (Figure \ref{turb}; Table \ref{param_turb} summarizes the parameters
used). The lower panel of Fig.\,\ref{turb} shows the residue of observed minus calculated
spectrum. Some remaining features may result from observational uncertainties (indicated
as dashed parts), for example the dip around 2.3225\,$\mu$m and the region around
2.314\,$\mu$m, whereas the features at 2.3253 and 2.3084\,$\mu$m seem to be yet
unidentified emission lines.

For a distance to MWC\,349 of 1.2\,kpc we find a CO emission area projected on the sky 
of $A_{\rm CO}\simeq 1.36\cdot 10^{26}$\,cm$^{2}$ and a mass of $M_{\rm CO}\simeq 3.2
\cdot 10^{24}$\,g. Our $A_{\rm CO}$ is about 5 times larger than the value of $2.5\cdot
10^{25}$\,cm$^{2}$ found by Geballe \& Persson (\cite{geballe}) who could only derive
a lower limit.


\subsection{CO bands from a Keplerian rotating disk}\label{disk}

\begin{figure}[htb]
\resizebox{\hsize}{!}{\includegraphics{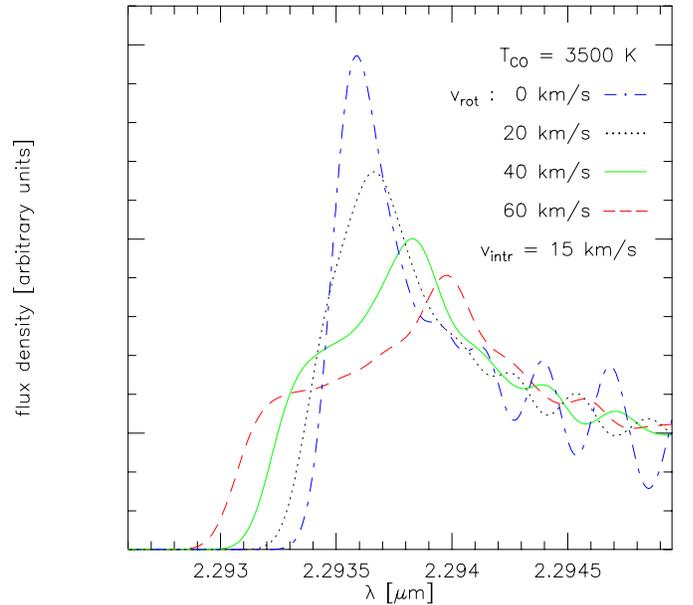}}
\caption{
Optically thin CO band emission from rings of constant rotational velocity
($v_{\rm rot} = 60,\,40,\,20$ and 0\,km/s) and for an inclination angle of $i =
0\degr$. The spectra are smoothed to the observational resolution of 15\,km/s.}
\label{rot}
\end{figure}

We investigate whether the CO bands in MWC\,349 can alternatively be explained
by emission from the circumstellar disk.  For demonstration, we first calculate
the optically thin emission from an infinitesimally narrow ring.  For Kepler
rotation, the orbital velocity is
\begin{equation}\label{kepler}
v_{\rm rot} = \sqrt{GM_*/r}\,.
\end{equation}
and the line of sight velocity, $v_{\rm los}$, of a gas element in the ring
\begin{equation}
v_{\rm los} = v_{\rm rot}\cdot\cos \theta\cdot\cos i~.
\end{equation}
Here $G$ denotes the gravitational constant, $r$ the ring radius, $M_{*}$ the
stellar mass, $i$ the inclination angle of the ring, and $\theta$ the azimuthal
angle.  The resulting profile of the CO band head smoothed to the observational
resolution of 15\,km/s shows a characteristic shoulder at short and a maximum at
long wavelengths relative to an unbroadened band head (see Fig.\,\ref{rot}).
The distance between shoulder and maximum decreases with decreasing orbital
velocity.  The CO profile of a rotating ring is evidently very different from
what is observed in MWC\,349.

We next calculate the emission from a whole disk viewed nearly edge-on, 
using the radiative transfer of Eq.\,(\ref{line_intens}). We
assume power laws for the temperature and surface density,
\begin{equation}\label{temp}
T(r) = T_{0} \cdot \left(\frac{r}{r_{i}}\right)^{-p} = T_{0} \cdot \left(\frac{v}
{v_{i}}\right)^{2p}
\end{equation}
and
\begin{equation}\label{sigma}
\Sigma(r) = \Sigma_{0} \cdot \left(\frac{r}{r_{i}}\right)^{-q} = \Sigma_{0} \cdot
\left(\frac{v}{v_{i}}\right)^{2q}~.
\end{equation}
$r_i$ is the inner radius of the Kepler disk where the rotational velocity has
its maximum value $v_{i}$.  We fix the inner radius by putting $v_{i} =
60$\,km/s (see Figure \ref{v_turb}).  Further, we assume $T_0 = 5000$\,K, which
is the dissociation limit of CO molecules, and an inclination angle of
$10\degr$.  The radius $r_{\rm out}$, out to which the CO (2$\to$1) band is
thermally excited, is a free parameter; the velocity there equals $v_{\rm 
out}$. For these disk calculations we surmize that the CO bands at 2 
micron are for radii greater than $r_{\rm out}$ only subthermally excited due 
to lower densities. Therefore their contribution to the overall spectrum is 
probably small. The CO disk is geometrically flat so that along each line of 
sight through the inclined disk the density and temperature are constant. The 
total flux from the disk is obtained from a straight forward numerical 
integration.

\begin{table}[htb]
\caption{Parameters for the fit of the CO disk calculations shown in
Fig.\,\protect{\ref{disk_fit}}. Fixed parameters are the mass of 26\,$M_{\sun}$, $T_{0}=
5000$\,K, $p = 0.5$, $q = 1.5$, and $v_i = 60$\,km/s which leads to $r_i\simeq
6.4$\,AU.}
\begin{tabular}{ccccc}
\hline
$v_{\rm out}$ & $r_{\rm out}$ & $T(r_{\rm out})$ & $N_{0}$ & $N(r_{\rm out})$ \\
$[$km/s] & [AU] & [K] & [cm$^{-2}$] & [cm$^{-2}$] \\
\hline
40 & 14.4 & 3333 & $8.5\cdot 10^{16}$ & $2.5\cdot 10^{16}$ \\
30 & 25.6 & 2500 & $5.8\cdot 10^{16}$ & $7.3\cdot 10^{15}$ \\
\hline
\end{tabular}
\label{disk_param}
\end{table}

\begin{figure}[htb]
\resizebox{\hsize}{!}{\includegraphics{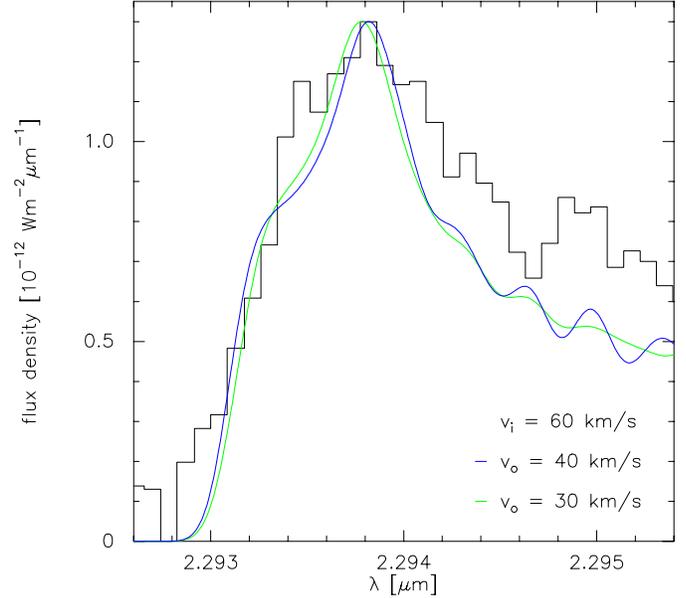}}
\caption{
Best fit to the observations (histogram) by Kepler disk.  Model parameters from
Table \protect{\ref{disk_param}}.  }
\label{disk_fit}
\end{figure}

By a systematic variation of the free parameters ($\Sigma_0, v_{\rm out}, p,q$),
we find that the shape of the spectrum is strongly influenced by the column
density at the inner edge and by the size of the disk, whereas the exponents $p$
and $q$ play a minor role; we therefore fix them following Hayashi
(\cite{hayashi}) to $p=0.5$ and $q=1.5$. Similar numbers are derived for the
dust disk (Kraus et al.~\cite{kraus}).

The disk models then contain only two free parameters: the outer radius, $r_{\rm
out}$, and the column density at the inner edge, $N_0$.  To explain the observed
band head profile with a rotating disk, one needs a large spread in the
rotational velocity because otherwise one cannot surpress the shoulders seen in
Fig. \ref{rot}.  The large velocity spread implies a large emitting area, and 
therefore a low column density, so that the emission becomes very optically
thin ($\tau\simeq 10^{-4}$), even at the inner edge of the CO disk. The outer
velocity $v_{\rm out}$ must lie in the range between 30 and 40\,km/s. If
$v_{\rm out}$ were higher, the shoulder of the band head would become too broad,
if it were smaller, the profile would get too narrow. Our best models of a
rotating disk are presented in Fig.~\ref{disk_fit}. Considering the quality of
the data, they are marginally acceptable. 

We add that if one assumes dust and gas to be thermally decoupled, it is in
these models nevertheless possible that the strength of the CO band head amounts
to $f \sim 5$\,\% of the continuum, as observed, although the CO emission is very optically
thin. As the dust emits like a blackbody, the equation $f\,B_{\nu}(T_{d}) =
\tau_{\nu}\,B_{\nu}(T_{\rm CO})$ is approximately fulfilled for $T_{d}\sim
1000$\,K, $T_{\rm CO} \sim 4000$\,K and $\tau\sim 10^{-4}$.

\section{Discussion}

\begin{figure}[b!]
\resizebox{\hsize}{!}{\includegraphics{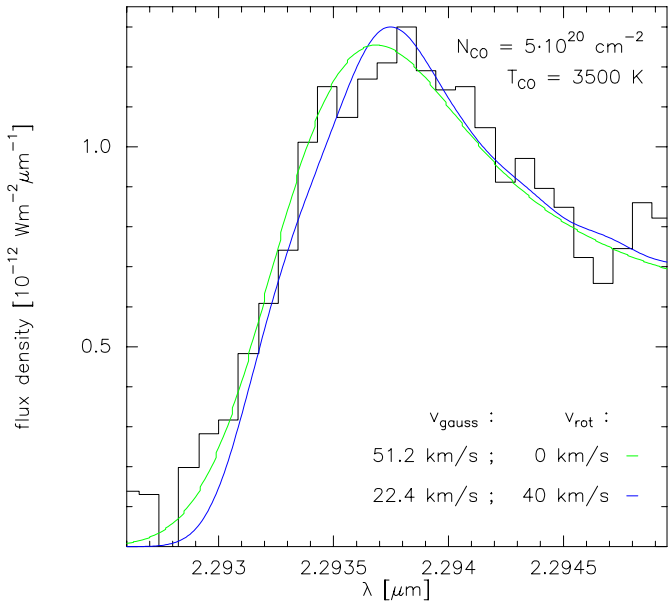}}
\caption{
Best fits to the observed $2\rightarrow 0$ band head. The grey line represents
the pure gaussian fit of Figure \protect{\ref{turb}} and the black line shows
the combination of a rotational component and a gaussian velocity component
whose velocities are determined according to Eq.\,\protect{\ref{l_min}}.}
\label{fit}
\end{figure}

Mathematically, there is also the possibility to combine the wind model of
Fig.~11 (solid line), which is based on a gaussian profile of $\sim 50$\,km/s
half width, with a Kepler rotation as long as $v_{\rm rot} \le 40$\,km/s. This
is shown in Figure \ref{fit}, where a rotational component with $v_{\rm rot} =
40$\,km/s (black curve) is added to the gaussian velocity profile. Otherwise
the basic parameters are the same as for the wind (LTE; CO column density $\sim
5\cdot 10^{20}$\,cm$^{-2}$; $T_{\rm CO} = 3500\ldots 4000$\,K).

\begin{figure*}[htb]
\psfig{file=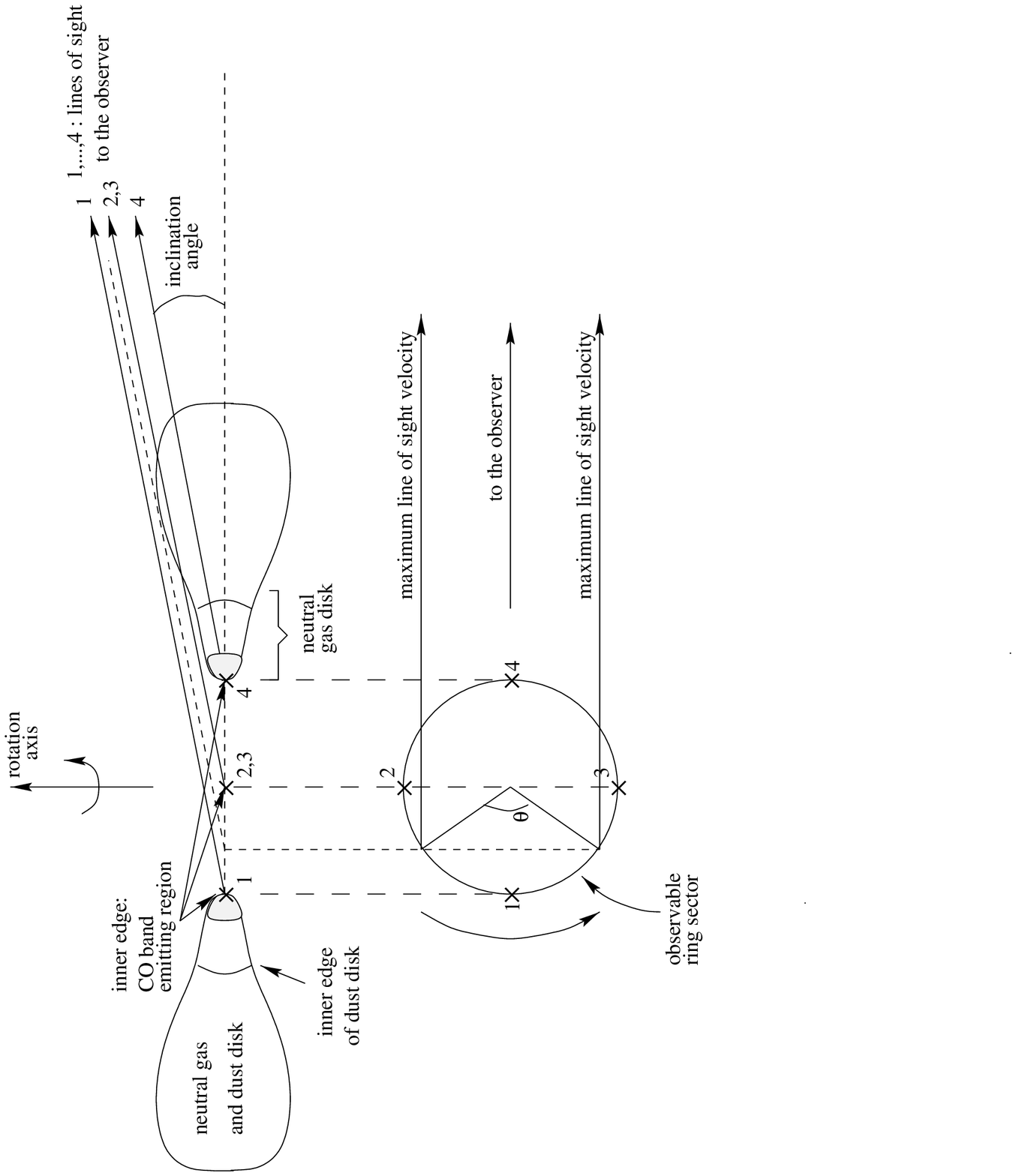,bbllx=45pt,bblly=101pt,bburx=408pt,bbury=694pt,scale=80,rotate=90,clip=}
\caption{
Sketch of the disk around MWC\,349 seen edge on.  Because of the small
inclination angle, only emission from sector 1 where the radial velocity is
small will reach the observer.  Light from other places, for example, from the
points marked 2, 3, and 4 is absorbed by the bulge.
}
\label{skizze4}
\end{figure*}

Although the fit is again very good, this scenario meets the following
difficulties: If the rotational velocity $v_{\rm rot} \le 40$\,km/s, the stellar
mass of 26\,M$_\odot$ implies a distance greater than 14\,AU.  If the CO gas is
coupled to the dust, it should have the same temperature of $\sim 950$\,K as the
dust grains at that distance (Kraus et al. \cite{kraus}), which is much too low.
If CO is located above the disk and thermally decoupled from the dust, it might
achieve the required temperature of $\sim 3500$\,K by radiative heating.  But
the CO column density of $5\cdot 10^{20}$\,cm$^{-2}$ then implies a gas surface
density which is at least an order of magnitude greater than what follows from
the dust disk (adopting standard conversion factors $\Sigma_{\rm dust} :
\Sigma_{\rm H} \simeq 10^{-2}$, $N_{\rm CO} : N_{\rm H} \simeq 10^{-4}$).

The scenario of a pure wind is very vague in its details, but gives a remarkably
good fit. We might think of dense molecular blobs emanating from the disk. The
total emission area of the blobs is $\sim 0.6$\,AU$^{2}$.

\begin{figure}[htb]
\resizebox{\hsize}{!}{\includegraphics{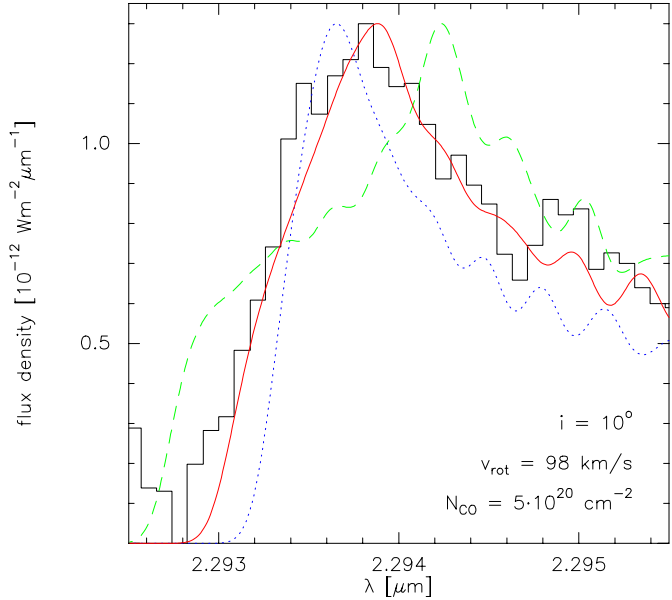}}
\caption{Fit to the observed $2\rightarrow 0$ band head for ring segments with a
Keplerian velocity of $v_{\rm rot} \simeq 100$\,km/s. The geometry is depicted
in Fig.\,17. The angles and the corresponding velocity ranges are given in
Table \protect{\ref{seg_param}}. The best fit is shown by the solid line.}
\label{segment}
\end{figure}

\begin{table}[htb]
\caption{Angles $\theta$ and velocity ranges $v_{\rm rot,los}$ for the
calculated sectors shown in Figure \protect{\ref{segment}}. The inclination
angle was taken as $10\degr$ and the rotational velocity is 98\,km/s.}
\begin{tabular}{r|rc}
\hline
 & $\theta [\degr]$ & $v_{\rm rot,los}$[km/s] \\
\hline
dashed line & 180 & $-96.5 \dots 96.5$ \\
solid line  &  75 & $-59.0 \dots 59.0$ \\
dotted line &  33 & $-27.6 \dots 27.6$ \\
\hline
\end{tabular}
\label{seg_param}
\end{table}

But there is still another alternative with $N_{\rm CO} \sim 5 \cdot 
10^{20}$\,cm$^{-2}$, which we discuss now. If the CO band emission arises in 
the disk, it must come from inside the evaporation radius of the dust, $r_{\rm 
evap}$, because of its high temperature of $\sim 3500$\,K. According to the 
Kraus et al.~model of the dust disk, the distance is then smaller than 2.88\,AU 
according to a rotational velocity greater than 90\,km/s.  Let us assume
$v_{\rm rot} \simeq 100$\,km/s. To avoid the problems with the line profile in 
case of such a high $v_{\rm rot}$, as discussed in subsection \ref{disk} and
demonstrated in Figure \ref{rot}, we propose a configuration as depicted in
Figure \ref{skizze4}. The disk has a bulge and because of its small inclination
angle $i\sim 10\degr$, this bulge blocks the light from the inner edge of the 
disk on the near side to the observer. We therefore receive emission only from 
sector 1 in Figure \ref{skizze4} where the {\it radial} velocity is low.
In Figure \ref{segment} we show computations of the flux received from sector 1
for different opening angles $\theta$. The ranges of radial velocities in the
sectors are summarized in Table \ref{seg_param}.

\section{Conclusions}

We present low and high resolution spectra in the near infrared of the B[e]-star
MWC\,349. The wavelength interval 2.285--2.342\,$\mu$m contains mainly
emission of the first overtone bands of the CO molecule as well as the Pfund
series of atomic hydrogen. From modeling the Pfund lines under the assumption
that they are optically thin, we find that they come from the inner part of the
\ion{H}{ii} region around MWC\,349 and are in LTE. The hydrogen line profiles 
are approximately gaussian and their width indicates a wind velocity of $\sim
50$\,km/s.

The CO molecules are at a temperature of 3500--4000\,K, the population of the
vibrational levels is close to LTE. The width of the $2\rightarrow 0$ band head
indicates a velocity broadening of the order of 50--60\,km/s, depending 
on the broadening mechanism, wind or rotation. The location of the hot CO gas producing the 
band emission could not be clarified, but we were able to propose several scenarios.

Because the CO as well as the Pfund lines can be fitted with a gaussian velocity
component of $\sim 50$\,km/s, the first possibility is that the CO bands arise
in the windy transition layer between the disk and the HII region, presumably 
in dense clumps (see Fig.\,13). The band head has in this case an optical depth 
of order unity ($N_{\rm CO}\simeq 5\cdot 10^{20}$\,cm$^{-2}$). 

Alternatively, the CO might form in a thin layer above the dust disk at radii
between 6.5 and 15\,AU. In this scenario, the CO emission is optically
very thin ($N_{\rm CO}\simeq 6\cdot 10^{16}$\,cm$^{-2}$; $\tau\simeq 10^{-4}$),
and the gas is thermally decoupled from the dust. As we could only produce fits 
of mediocre quality (see Fig.\,15), due to the characteristic line profiles
under Keplerian rotation (see Figure \ref{rot}), this is not our favorite
configuration.

Finally, we suggest that the CO bands come from the inner edge of the 
circumstellar disk. Because of the required high temperature and column density,
the CO gas must be located inside the evaporation radius of the dust, i.e.~at
distances less than $\sim 3$\,AU from the star. Although this would imply a
high rotational velocity component which would usually lead to a characteristic 
shoulder in the band profile, which is not observed, the disk may well have a
bulge, as depicted in Figure \ref{skizze4}. This bulge absorbs the CO emission
from the near part of the rotating inner disk edge, so that the observer sees 
only a sector on the far side where the radial velocities are all smaller than 
$\sim 60$\,km/s. With such a geometrical configuration, a satisfactory fit is 
also possible (see Figure \ref{segment}).

\begin{acknowledgements}

We would like to thank Frank Shu, the referee, for his helpful comments

\end{acknowledgements}


\begin{thebibliography}{}

\bibitem[1991]{calvet}
         Calvet, N., Pati\~{n}o, A., Magris, C.G., D`Alessio, P., 1991,
         ApJ 380, 617
\bibitem[1989]{carr}
         Carr, J.S., 1989,
         ApJ 345, 522
\bibitem[1995]{carr95}
         Carr, J.S., 1995,
         Ap\,\&\,SS 224, 25
\bibitem[1993]{carr93}
         Carr, J.S., Tokunaga, A.T., Najita, J., Shu, F.H.,
         Glassgold, A.E., 1993,
         ApJ 411, L\,37
\bibitem[1993]{chandler}
         Chandler, C.J., Carlstrom, J.E., Scoville, N.Z.,
         Dent, W.R., Geballe, T.R., 1993,
         ApJ 412, L\,71
\bibitem[1995]{chandler95}
         Chandler, C.J., Carlstrom, J.E., Scoville, N.Z., 1995,
         ApJ 446, 793
\bibitem[1996]{chandra}
         Chandra, S., Maheshwari, V.U., Sharma, A.K., 1996,
         A\&AS 117, 557
\bibitem[1985]{cohen}
         Cohen, M., Bieging, J.H., Dreher, J.W., Welch, W.J., 1985,
         ApJ 292, 249
\bibitem[1932a]{dunham32a}
         Dunham, J. L., 1932a,
         Phys.Rev. 41, 713
\bibitem[1932b]{dunham32b}
         Dunham, J. L., 1932b,
         Phys.Rev. 41, 721
\bibitem[1991]{farrenq}
         Farrenq, R., Guelachvili, G., Sauval, A. J.,
         Grevesse, N., Farmer, C. B., 1991,
         J.Mol.Spectrosc. 149, 375
\bibitem[1987]{geballe}
         Geballe, T.R., Persson, S.E., 1987,
         ApJ 312, 29
\bibitem[1970]{geisel}
         Geisel, S.L., 1970,
         ApJ 161, L\,105
\bibitem[1996]{green}
         Green, T.P., Lada, C.J., 1996,
         ApJ 461, 345
\bibitem[1986]{hamann}
         Hamann, F., Simon, M., 1986,
         ApJ 311, 909
\bibitem[1988]{hamann88}
         Hamann, F., Simon, M., 1988,
         ApJ 327, 876
\bibitem[1981]{hayashi}
         Hayashi, C., 1981,
         Suppl.Prog.Theor.Phys. 70, 35
\bibitem[2000]{kraus00}
         Kraus, M., 2000, PhD-Thesis, University of Bonn
\bibitem[2000]{kraus}
         Kraus, M., Kr\"ugel, E., Hengel, C., Thum, C., 2000
         {\it in preparation}
\bibitem[1998]{lamers}
         Lamers, H.J.G.L.M., Zickgraf, F.-J., de Winter, D.,
         Houziaux, L., Zorec, J., 1998,
         A\&A 340, 117
\bibitem[1986]{leinert}
         Leinert, C., 1986,
         A\&A 155, L\,6
\bibitem[1983]{mariotti}
         Mariotti, J.M., Chelli, A., Foy, R., L\'ena, P., Sibille, F.,
         Tchountonov, G., 1983,
         A\&A 120, 237
\bibitem[1997]{martin}
         Martin, S.C., 1997,
         ApJ 478, L\,33
\bibitem[1989]{martin-pintado}
         Mart\'{\i}n-Pintado, J., Bachiller, R., Thum, C.,
         Walmsley, M., 1989,
         A\&A 215, L\,13
\bibitem[1935]{menzel}
         Menzel, D.H., Pekeris, C.L., 1935,
         MNRAS 96, 77
\bibitem[1996]{najita}
         Najita, J., Carr, J.S., Glassgold, A.E., Shu, F.H.,
         Tokunaga, A.T., 1996,
         ApJ, 462, 919
\bibitem[1994]{rodriguez}
         Rodr\'{\i}guez, L.F., Bastian, T.S., 1994,
         ApJ 428, 324
\bibitem[1979]{scoville}
         Scoville, N., Hall, D.N.B., Kleinmann, S.G., Ridgeway, S.T., 1979,
         ApJ 232, L\,121
\bibitem[1997]{smith}
         Smith, H.A., Strelnitski, V., Miles, J.W., Kelly, D.M.,
         Lacy, J.H., 1997,
         AJ 114, 2658
\bibitem[1995]{storey}
         Storey, P.J., Hummer, D.G., 1995,
         MNRAS 272, 41
\bibitem[1996]{strelnitski96}
         Strelnitski, V.S., Haas, M.R., Smith, H.A., Erickson, E.F.,
         Colgan, S.W.J., Hollenbach, D.J., 1996a,
         Sci 272, 1459
\bibitem[1992]{thum}
         Thum, C., Mart\'{\i}n-Pintado, J., Bachiller, R., 1992,
         A\&A 256, 507
\bibitem[1998]{thum98}
         Thum, C., Mart\'{\i}n-Pintado, J., Quirrenbach, A.,
         Matthews, H.E., 1998,
         A\&A 333, L\,63
\bibitem[1994]{thum94}
         Thum, C., Matthews, H.E., Mart\'{\i}n-Pintado, J., Serabyn, E.,
         Planesas, P., Bachiller, R., 1994b,
         A\&A 283, 582
\bibitem[1995]{thum95}
         Thum, C., Strelnitski, V.S., Mart\'{\i}n-Pintado, J.,
         Matthews, H.E., Smith, H.A., 1995,
         A\&A 300, 843
\bibitem[1985]{white}
         White, R.L., Becker, R.H., 1985,
         ApJ 297, 677


\end{thebibliography}
\end{document}